# Dilemma in All-optical Characterization of Single-layer NiI$_2$ Multiferroics

*Yucheng Jiang[1,2,3], Yangliu Wu[4], Jinlei Zhang[2], Jingxuan Wei[1], Bo Peng[4,\*] and Cheng-Wei Qiu[1,\*]*

[1]Department of Electrical and Computer Engineering, National University of Singapore, 4 Engineering Drive 3, Singapore 117583, Singapore

[2]Jiangsu Key Laboratory of Micro and Nano Heat Fluid Flow Technology and Energy Application, School of Physical Science and Technology, Suzhou University of Science and Technology, Suzhou, 215009, China

[3]Science, Mathematics and Technology (SMT), Singapore University of Technology and Design (SUTD), 8 Somapah Road, Singapore 487372, Singapore

[4]National Engineering Research Center of Electromagnetic Radiation Control Materials, School of Electronic Science and Engineering, University of Electronic Science and Technology of China, Chengdu 611731, China

\*e-mail: bo_peng@uestc.edu.cn and chengwei.qiu@nus.edu.sg

The search for two-dimensional multiferroic materials is an exciting yet challenging endeavor. Recently, Song et al. reported the exciting discovery of type-II multiferroic order in an antiferromagnetic (AFM) NiI$_2$ single layer, and concluded the ferroelectric (FE) polarization induced by a helical magnetic order[1]. Their finding was presented from all-optical experimental evidence, based on the methods of second harmonic generation (SHG) and linear dichroism (LD). However, the all-optical characterizations alone cannot serve as unanimous evidence of the "FE" existence, particularly in conjunction with magnetic orders in a single-layer multiferroic. A main argument is that their observed SHG and LD signals may just be manifested from the magnetic-order induced breaking of spatial-inversion or time-reversal symmetry instead of FE order, and thus cannot endorse the finding of a single-layer multiferroic.

Accurate measurements of magnetic and electric orders are very essential to make a reliable judgement of a multiferroic. With the device's thickness approaching atomic scale, its magnetic or electrical signals will become too weak to be collected by traditional measurements[2]. Therefore, advanced all-optical probing methods have been deployed to characterize the magnetism and ferroelectricity, especially for a single layer, such as magneto-optical Kerr effect, reflective magnetic circular dichroism (RMCD), SHG and LD [2,3]. Among them, SHG and LD represent two typical optical approaches to investigate the domain structure and phase transition of ferroelectrics[4,5]. As indirect characterizations of ferroelectricity, their signals were thought to reflect the breaking of lattice inversion symmetry. To date, for a single ferroic, optical methods have made great success in the exploration of ferromagnetism or ferroelectricity in two-dimensional systems[2,4,6]. However, all-optical methods are unreliable to make a judgement of single-layer multiferroics, especially for the antiferromagnets[7-9]. In this case, the SHG and LD responses are not necessarily associated with FE polarization, but reflect the magnetic-order induced breaking of inversion symmetry. Without electric and magnetic fields applied, the all-optical methods cannot scrutinize the origin of optical signals.

In the original paper,[1] the SHG and LD data were interpreted as solid evidence for ferroelectricity in the $NiI_2$ single layer, in conjunction with the observation of helical magnetic ($T_{N,2}$) phase transition. It was an important experimental basis to identify the single-layer $NiI_2$ as a Type-II multiferroic. However, several previous works have revealed the inconvenient truth that the magnetic order can also lead to SHG and LD signals directly by breaking an inversion symmetry (see Extended Data Fig. 1)[7,8], without the need of inducing an FE polarization. For example, an AFM $CrI_3$ bilayer exhibits a giant electric-dipole SHG, despite its centrosymmetric crystal structure[7]. The observed SHG domains tally with AFM domains, both of which show synchronous responses to the external magnetic field (see Figs. 1 f-g in Ref.[7]). That means the SHG actually reflects the information of magnetic structure instead of electric structure. SHG signals originates from the AFM-order induced breaking of spatial-inversion and time-reversal symmetries. Another typical example is shown in Ref.[8] (see Extended Data Fig. 1c). In the $FePS_3$ crystal, a sharp onset in LD signature is observed at 118 K, at which AFM phase transition occurs too. It is justified that the SHG and LD data cannot sufficiently prove the existence of ferroelectricity, particularly in the presence of antiferromagnetism. For the $NiI_2$ single layer, the authors provided the data of polarization domains and phase transition through all-optical methods and conventional theoretical analysis[1]. But, magnetic orders not only break the inversion symmetry, but also induce rotation of polarization plane due to Faraday effect. Strictly, LD, SHG and polarized microscopy images cannot sufficiently validate FE polarization in the presence of magnetic orders. The coexistence of FE and magnetic orders complicates the situation, causing that it is impossible to distinguish FE polarization from magnetic orders through all-optical results.

In Fig. 1a, we analyze the effect of magnetic order on the symmetry of $NiI_2$. After an operation of spatial-inversion ($r \rightarrow -r$) operation or a time-reversal ($t \rightarrow -t$), $NiI_2$ is converted to a different magnetic state. The helical magnetic structures along the propagation vector $Q_x$ break both spatial-inversion and time-reversal symmetries. In this case, one cannot exclude the possibility that the magnetic order causes SHG and LD signals. The resulting optical domains may correspond to the magnetic domains instead of FE domains in $NiI_2$. To compare magnetic domains with optical domains, we performed the RMCD and LD measurements in few-layer $NiI_2$ with different magnetic fields applied as shown in Figs. 1 b and c. It is well known that the RMCD map can detect the magnetic domains. Good consistency has been shown between RMCD and LD maps. With an external magnetic field applied, the LD domains change synchronously with the switching of magnetic domains. Therefore, the observed LD signals are more likely to originate from the magnetic structure, similar to the cases of $CrI_3$ bilayer and few-layer $FePS_3$.

In general, SHG and LD can be widely captured in a variety of materials with drastically different origins (see Table 1)[4,10-13]. For non-ferroic and FE materials, SHG and LD can be attributed to non-centrosymmetric lattice structures. For centrosymmetric AFM materials, they result from breaking inversion symmetry induced by AFM order. Without electric or magnetic fields applied, it is difficult to clarify the origin of optical signals only by using all-optical methods.

In our opinion, the $NiI_2$ single layer may be a candidate as a two-dimensional multiferroic, but requires more direct evidence apart from those all-optical results. The given experimental data are not sufficient to support the authors' conclusion[1]. It calls for more direct magnetic and electric measurements to confirm the multiferroic properties of $NiI_2$ single layer. The original paper lacks the magnetic- and electric-field control of SHG or LD domains, making it impossible to determine whether optical signals result from FE polarization or magnetic order. Next, the authors should probe the FE domains by using a direct electrical method like piezoresponse force microscopy. If possible, it would be better if they compare the FE domains with SHG domains directly by mapping the experimental data. It is therefore believed that the discovery of multiferroic single layer is questionable based on the current evidence.

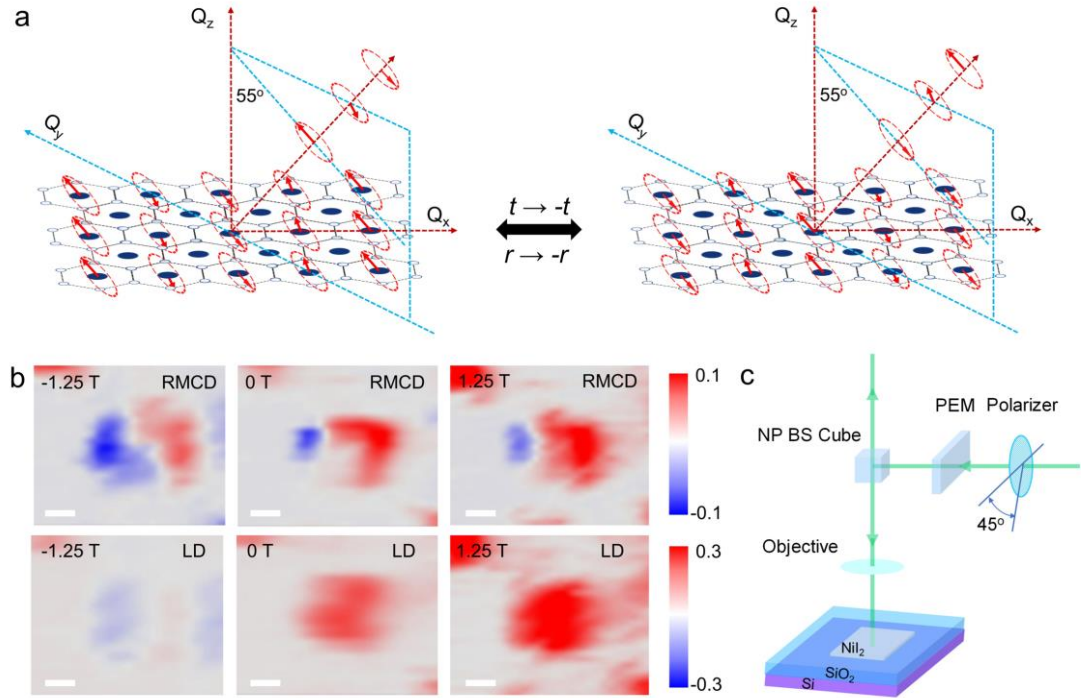

**Fig. 1| Magnetic order and optical domains of NiI$_2$. a,** Schematic of the helical magnetic structure of NiI$_2$, showing a spatial-inversion ($r \rightarrow -r$) or time-reversal ($t \rightarrow -t$) operation. **b,** RMCD and LD intensity images of few-layer NiI$_2$ under different magnetic fields at 11 K for comparison. The scale bar is 1 μm. **c,** Schematic of RMCD and LD measurements for a NiI$_2$ flake (see Methods).

**Table 1| Classification of the materials showing the SHG and LD signals.**

| | Non ferroic | Single ferroic | | Multiferroic |
|---|---|---|---|---|
| Type | Non FE or AFM | FE | AFM | FE+AFM |
| Breaking of inversion symmetry | | | | |
| SHG | α-SiO$_2$[4] | BaTiO$_3$[10] | CrI$_3$ bilayer[7] | MnWO$_4$[12] |
| LD | Black phosphorus[11] | In$_2$Se$_3$[5] | FePS$_3$[8] | Pb(TiO)Cu$_4$(PO$_4$)$_4$[13] |

**Methods**

A 532 nm laser was coupled to the Witec Raman system with closed cycle superconducting magnet and He optical cryostat. The light of ~0.6 μW was modulated by photoelastic modulator and focused onto samples by a long working distance objective (50×, NA = 0.45). The reflected beams were collected by the same objective, passed through a non-polarizing beamsplitter cube into a photodetector (see Fig. 1c). For RMCD measurements, the retardance is λ/4 and PEM frequency is 50 kHz. For LD measurements, the retardance and PEM frequency is doubled. The detailed experimental and measurements by magneto-optical-electric joint-measurement scanning imaging system (MOEJSI) are shown in the following parts.

Data availability

All data are available from the corresponding author upon reasonable request.

**Acknowledgements** C.W.Q. would like to acknowledge the support from National Research Foundation Singapore (CRP26-2021-0004). B.P. thanks the financial support from National Science Foundation of China (52021001).

**Author contributions** C.W.Q. conceived this work. Y.C.J. and B.P. wrote the manuscript with inputs of C.W.Q. and J.L.Z. Y.L.W. and B.P. performed the RMCD and LD experiments. All of authors contributed to the discussion.

**Competing interests** The authors declare no competing interests.

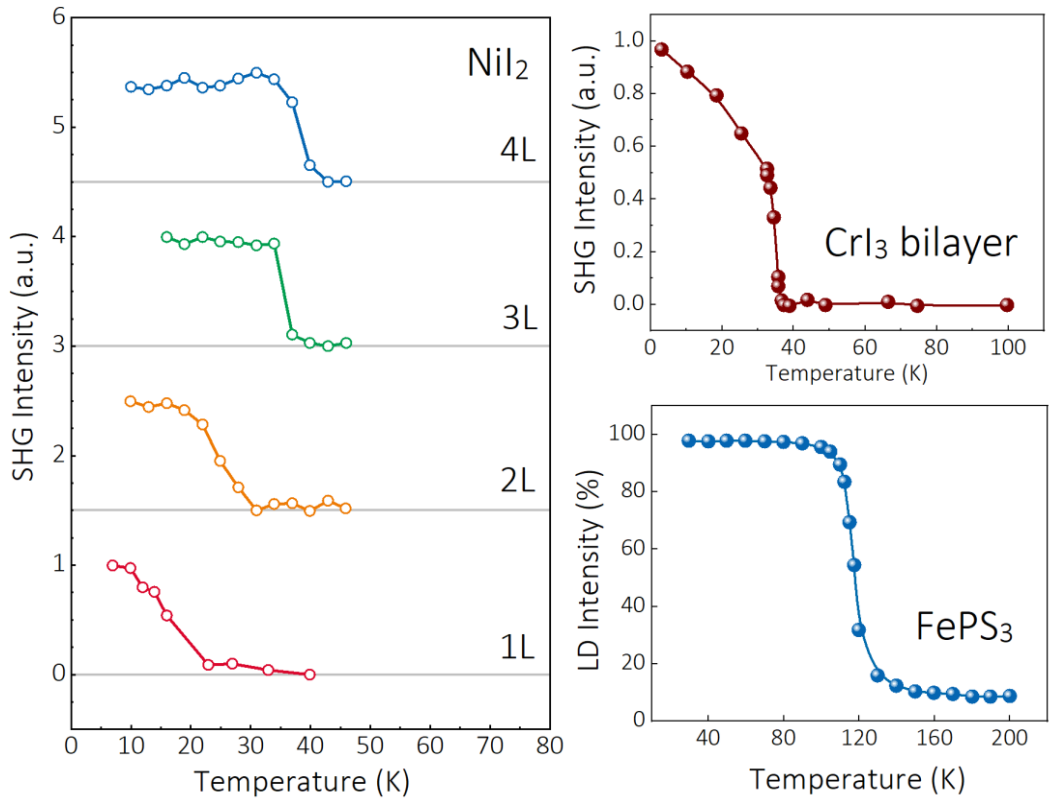

**Extended data Fig. 1| Reproduced SHG and LD responses of NiI$_2$ (Ref.[1]), CrI$_3$ (Ref.[7]) and FePS$_3$ (Ref.[8]).** Temperature dependence of SHG intensity **a,** for one- to four-layer NiI$_2$ and **b,** for CrI$_3$ bilayer. **c,** LD intensity as a function of temperature for a FePS$_3$ multilayer, showing a phase transition at 118 K. The data are collected from Ref.[1,7,8].

# Magneto-optical-electric joint-measurement scanning imaging system for identification of two-dimensional vdW multiferroic

*Yangliu Wu, Bo Peng*[*]

School of Electronic Science and Engineering, University of Electronic Science and Technology of China, Chengdu 611731, China

*e-mail: bo_peng@uestc.edu.cn

**Sample fabrication**

$NiI_2$ flakes were prepared by a normal and widely-used mechanical exfoliation method. The few-layer $NiI_2$ were exfoliated via PDMS films in a glovebox from $NiI_2$ bulk crystals, synthesized by chemical vapor transport method from elemental precursors with molar ratio Ni:I = 1:2. All exfoliated hBN, $NiI_2$ and graphene flakes were transferred onto pre-patterned Au electrodes on $SiO_2$/Si substrates one by one to create a heterostructure in glovebox and were further in-situ loaded into a microscopy optical cryostat for optical measurements in the glovebox. The whole process of $NiI_2$ sample fabrication and measurement were kept out of atmosphere[1,2].

**Optical measurements by the magneto-optical-electric joint-measurement scanning imaging system (MOEJSI)**

As an advanced imaging system, the magneto-optical-electric joint-measurement scanning imaging system (MOEJSI) brings spectroscopic techniques with unmatched spatial resolution to very low temperature, high magnetic field and high electric field measurements[3]. It was developed for investigating the magnetic and ferroelectric properties and their mutual control through magneto-optical-electric joint-measurements, besides Raman and photoluminescence features[4-7]. In particular, the RMCD loops and imaging, linear dichroism (LD) imaging and polarization-electric field hysteresis loop can be achieved when simultaneously applied high magnetic field (7 T) and electric field (100 V) at low temperature of 10 K.

The MOEJSI system is built based on a Witec alpha 300R Plus low-wavenumber confocal Raman imaging microscope with a spatial resolution reaching diffraction limit, which is integrated with a closed cycle superconducting magnet (7 T) with a room temperature bore and a closed cycle cryogen-free microscopy optical cryostat (10 K) with electronic transport measurement assemblies[8-10], as shown in Fig. 1a. This system achieves multi-field coupling of magnetic fields, electric fields and optical fields and realize high speed, sensitivity and resolution. For use in the room temperature bore superconducting magnet, an extended snout sample mount is specially designed for the cryogen-free microscopy optical cryostat (Fig. 1b). The microscopy optical cryostat directly anchored on the XY scanning stage of Witec Raman system for scanning imaging. The objective mounted in a lens tube and the extended snout enter into the room-temperature bore from top and bottom, respectively (Fig. 1b and 1c).

For RMCD and LD measurements, a free-space 532 nm laser of ~2 μW was linearly polarized at 45° to the photoelastic modulator (PEM) slow axis and sinusoidally phase-modulated by PEM (RMCD: 50 KHz; LD: 100 KHz), with a maximum retardance of λ/4 for RMCD and λ/2 for LD. The light was further reflected by a non-polarizing beamsplitter cube (R/T = 30/70) and then directly focused onto samples by a long working distance 50× objective (Zeiss, WD = 9.1 mm, NA = 0.55). The reflected beam which was collected by the same objective passed through the same non-polarizing beamsplitter cube and was detected by a photomultiplier (PMT). The PMT and PEM coupled with lock-in amplifier and Witec scanning imaging system (Fig. 2 and 3). The RMCD and LD maps as a function of magnetic field are shown in Fig. 4. With a magnetic field, the LD domains change synchronously with the switching of magnetic domains. Therefore, the observed LD signals originate from the magnetic structure, rather than ferroelectric polarization. Only the LD optical method can not identify the existence of ferroelectric polarization and is not reliable. The electric measurements, including *P-E*, *C-E* and *J-E* hysteresis loop must be done and mutual control of magnetic orders and ferroelectric polarization should be observed.

**Data availability**
All data are available from the corresponding author upon reasonable request.

**Acknowledgements** B.P. thanks the financial support from National Science Foundation of China (52021001, 62250073).

**Author contributions** B.P. conceived this work. Y.L.W. and B.P. performed the RMCD and LD experiments. All of authors contributed to the discussion.

**Competing interests** The authors declare no competing interests.

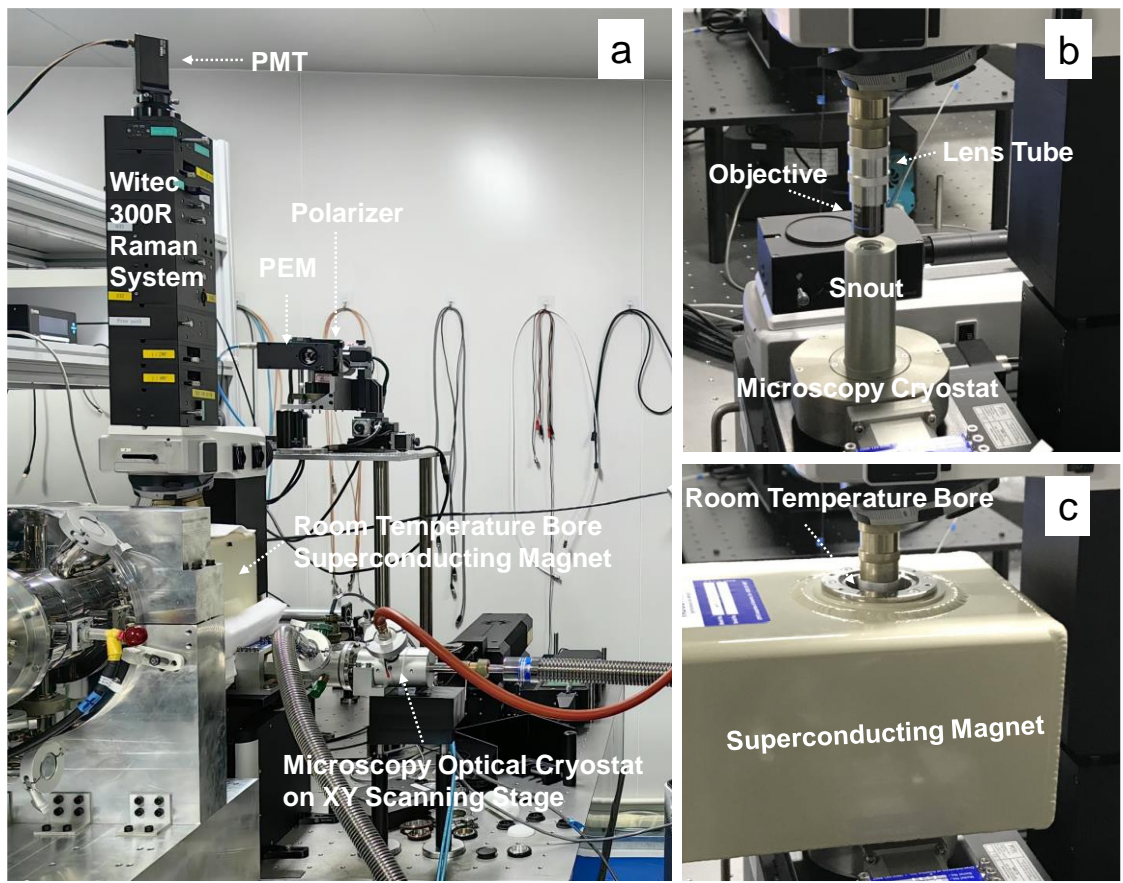

**Fig. 1 | The photograph of the magneto-optical-electric joint-measurement scanning imaging system (MOEJSI)**. **a**, Witec 300R Raman imaging microscope, closed cycle superconducting magnet (7 T) with room-temperature bore and cryogen-free microscopy optical cryostat (10 K) with extended snout sample mount are integrated together. **b**, An objective in an extended tube is coupled with the extended snout of the microscopy optical cryostat. **c**, The objective and the extended snout get into the room-temperature bore of superconducting magnet from top and bottom, respectively.

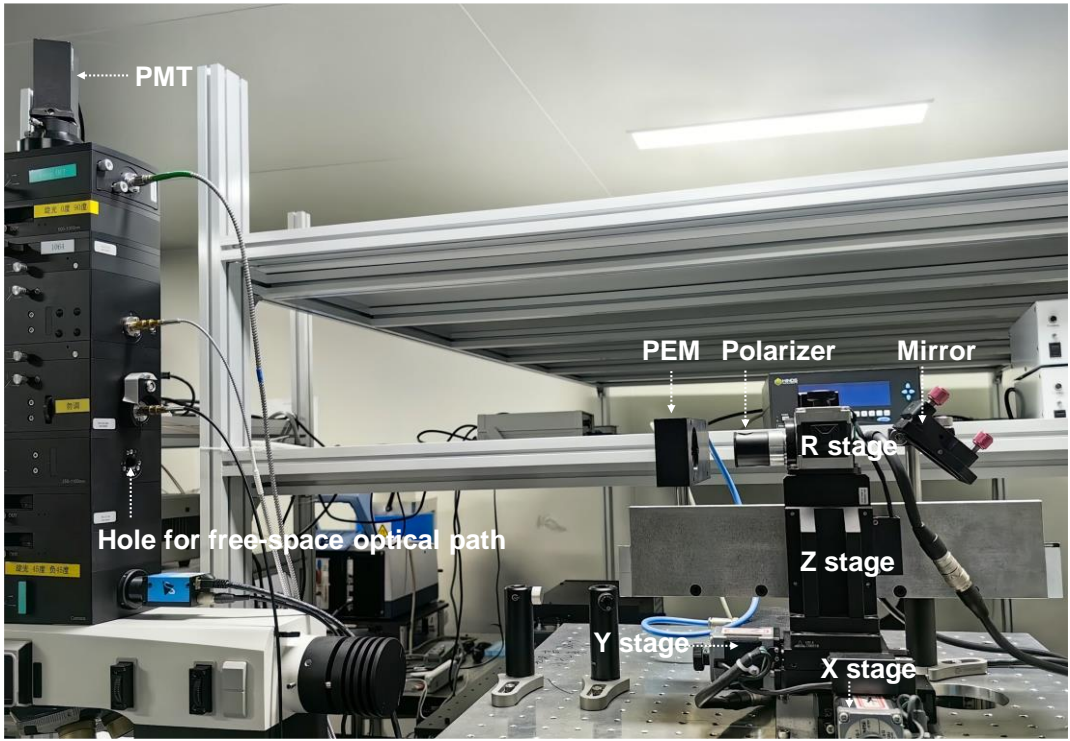

**Fig. 2 | The photograph of free-space RMCD and LD optical path.**

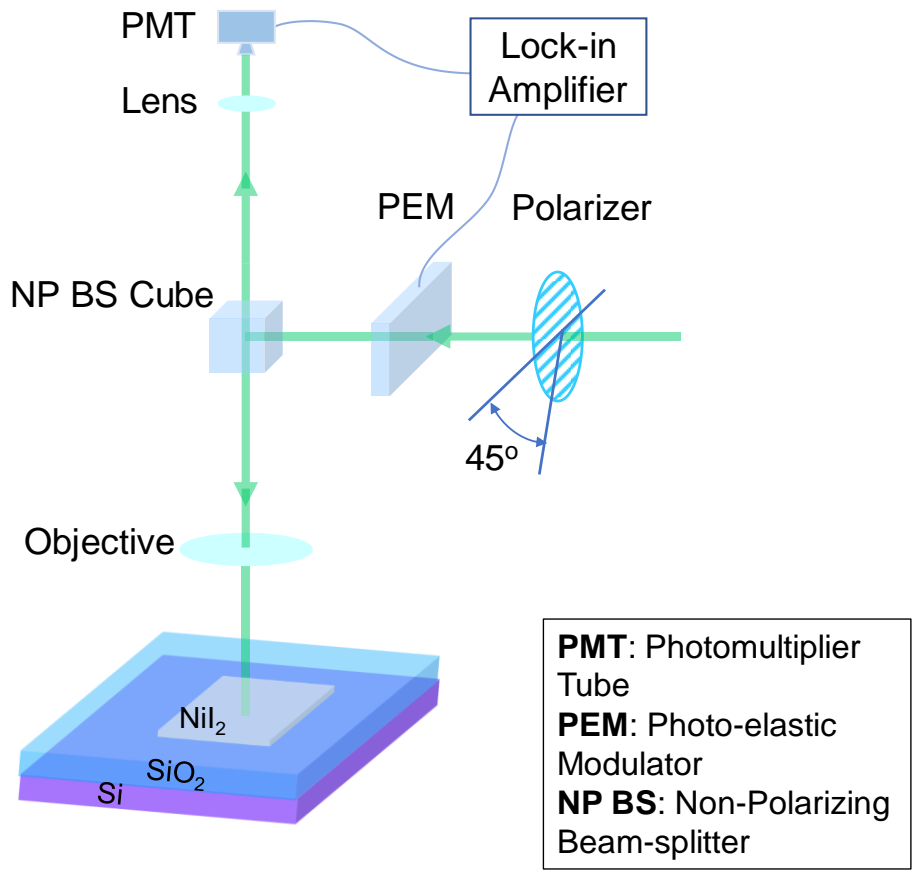

**Fig. 3 | Schematic of free-space RMCD and LD optical path.**

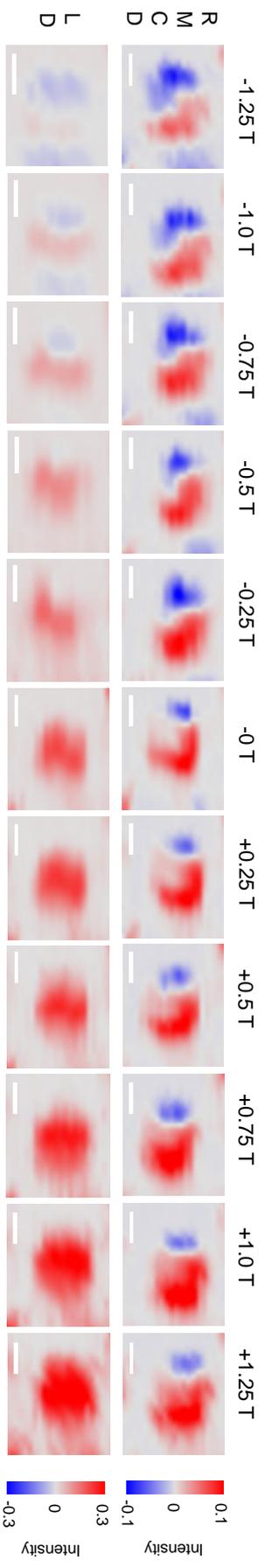

**Fig. 4 | RMCD and LD maps as a function of magnetic field. The scale bar is 1 μm.**